\documentclass[aps,pra,twocolumn,showpacs,groupedaddress,letterpaper]{revtex4-1}

\usepackage{graphicx}
\usepackage{subfigure}
\usepackage{color}
\usepackage{physymb}
\usepackage{enumitem}
\usepackage{mdframed}
\usepackage{framed}
\usepackage{epstopdf}

\begin{document}

\title{Validation and analysis of the coupled multiple response Colorado upper-division electrostatics (CUE) diagnostic}

%\keywords{physics education research, electrostatics, conceptual assessment, CUE, multiple-choice}

\author{Bethany R. Wilcox}
\affiliation{Department of Physics, University of Colorado, 390 UCB, Boulder, CO 80309}

\author{Steven J. Pollock}
\affiliation{Department of Physics, University of Colorado, 390 UCB, Boulder, CO 80309}

\begin{abstract}
Standardized conceptual assessment represents a widely-used tool for educational researchers interested in student learning within the standard undergraduate physics curriculum.  For example, these assessments are often used to measure student learning across educational contexts and instructional strategies.  However, to support the large-scale implementation often required for cross-institutional testing, it is necessary for these instruments to have question formats that facilitate easy grading.  Previously, we created a multiple-response version of an existing, validated, upper-division electrostatics diagnostic with the goal of increasing the instrument's potential for large-scale implementation.  Here, we report on the validity and reliability of this new version as an independent instrument.  These findings establish the validity of the multiple-response version as measured by multiple test statistics including item difficulty, item discrimination, and internal consistency.  Moreover, we demonstrate that the majority of student responses to the new version are internally consistent even when they are incorrect, and provide an example of how the new format can be used to gain insight into student difficulties with specific content in electrostatics.  
\end{abstract}

\maketitle

\section{\label{sec:introduction}Introduction \& Background}

One natural focus of the Physics Education Research (PER) community is on understanding and improving student learning in our physics courses.  Often, a critical component of this research is achieving valid measures of student learning, both before and after instruction.  Moreover, it is often important that these measures be standardized so that they can assess student learning across populations, learning environments, and instructional strategies.  Research-based conceptual assessments, such as the Force Concept Inventory (FCI) \cite{hestenes1992fci} and Brief Electricity and Magnetism Assessment (BEMA) \cite{ding2006bema}, are often used for this purpose.  After careful validation, these instruments provide a standardized measure of student understanding of specific physics content.  Previously, student performance on these assessments has been used to help motivate educational transformation efforts aimed at supporting increased student learning \cite{finkelstein2005tutorial} as well as by individual physics instructors to provide formative feedback on the effectiveness of their own instructional practices.  

A wide variety of conceptual assessments that target introductory physics content have been developed (see Ref.\ \cite{beichnerAssessment} for a list), and recently a smaller number of upper-division assessments have also been created \cite{wilcox2015assessment}.  Upper-division conceptual assessments are rarer in part because the more advanced physics content of these courses presents unique challenges including the necessary use of specialized language and sophisticated mathematics.  One example of an existing upper-division assessment is the Colorado Upper-division Electrostatics (CUE) diagnostic \cite{chasteen2012cue}.  The CUE was designed to target junior-level electrostatics content (i.e., chapters 1-6 of Griffiths \cite{griffiths1999em}).  Unlike its introductory counterparts, the questions on the CUE primarily have a free response, rather than multiple-choice, format in order to more effectively target students' ability to synthesize and generate responses. 

\begin{figure}
  \begin{minipage}{1\linewidth}
  \begin{framed}
    \vspace{-1mm}\flushleft {\bf Q5} - A charged, insulating solid sphere of radius $R$ with a uniform volume charge density $\rho_o$, with an off-center spherical cavity of radius $r$ carved out of it (see Figure).  \\ \vspace{1mm}Find $\vec{E}$ or $V$ at point P, a distance {\bf $4R$} from the sphere.\\
    \vspace{1mm} Select only one: {\bf The easiest method would be ...}

   \begin{minipage}{0.52\linewidth}
      \vspace{1mm}\flushleft
      A. Direct Integration\\
      B. Gauss's Law\\
      C. Separation of Variables\\
      D. Multipole Expansion\\
      E. Ampere's Law\\
      F. Method of Images\\
      G. Superposition\\
      H. None of these
   \end{minipage}
   \begin{minipage}{0.45\linewidth}    
      \begin{center}
        \includegraphics[width=24mm]{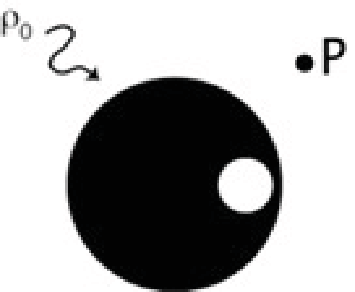}
      \end{center}
   \end{minipage}
     \flushleft{{\bf because ...}(select {\bf ALL} that support your choice)\\
      a. $\Box$ you can calculate $\vec{E}$ or $V$ using the integral form \\ \hspace{8mm}of Coulomb's Law\\
      b. $\Box$ the sphere will look like a dipole; approximate with \\ \hspace{8mm}$\vec{E}$ or $V$ for an ideal dipole\\
      c. $\Box$ $\vec{E}$ or $V$ outside a uniform sphere is the same as\\ \hspace{8mm}from a point charge at the center\\
      d. $\Box$ the location of the cavity doesn't matter, you just \\ \hspace{8mm}need $Q_{enclosed}$ to calculate $\vec{E}$\\
      e. $\Box$ you can treat this as two uniform spheres, one with \\ \hspace{8mm}charge density $\rho_o$ and one with charge density $-\rho_o$\\
      f. $\Box$ this will be the same as a uniform sphere with total \\ \hspace{8mm}charge $\frac{4}{3}\pi(R^3-r^3)\rho_o$\\
      g. $\Box$ electric fields from multiple sources can be \\ \hspace{8mm}combined through a vector sum\\
      h. $\Box$ $\nabla^2 V = 0$ outside the cube and you can solve for V \\ \hspace{8mm}using Fourier Series} \vspace{-4pt}
  \end{framed}
  \end{minipage}
\caption{A sample item from the CMR CUE.  The prompt has been paraphrased; see Ref.\ \cite{wilcox2014cmr} the full prompt and a detailed discussion of question development and scoring.} \label{fig:exampleCMR}
\end{figure}

To date, the CUE has been used productively to assess student learning for a number of semesters at multiple institutions \cite{chasteen2012cue, wilcox2015assessment}.  The CUE and its associated scoring rubric have been shown to be both valid and reliable for use with this population of upper-division physics students.  The assessment has also demonstrated a sensitivity to different types of instruction (e.g., interactive engagement vs. lecture only).  Recently, in response to the challenges inherent in scoring this type of free-response instrument on a large-scale, we crafted a new version of the assessment known as the Coupled Multiple-Response (CMR) CUE.  The CMR format utilizes a tiered multiple-response format in which students can select multiple response options and receive credit based on both the accuracy and consistency of their responses.  An example of this multiple-response format is given in Fig.\ \ref{fig:exampleCMR}.  The development of the CMR CUE is described in detail in a companion publication \cite{wilcox2014cmr}.  Briefly, the CMR version was created using student responses to the free-response version to construct distractors.  As part of its development, the new version was reviewed by 9 physics faculty and post-docs to ensure that the physics content was clearly and correctly expressed.  The instrument was also administered to 13 students in a think-aloud interview setting to ensure that students were interpreting the questions and distractors as intended.  

The CMR CUE is scored with an electronic scoring spreadsheet, thus preserving the fast, objective grading of a standard multiple-choice instrument.  This spreadsheet utilizes a nuanced grading scheme designed to match the scoring of the original free-response version \cite{wilcox2014cmr}.  In this scheme, students responding to items like that in Fig.\ \ref{fig:exampleCMR} are awarded points for selecting the easiest method and the correct reasoning elements.  They can also receive points for selecting methods that are possible, but not easy, or for selecting reasoning elements that are consistent with their choice of method.  Students can also lose reasoning points if they select reasoning elements that are inconsistent with their choice of method.  We also explored simpler grading schemes that, for example, did not offer partial credit or detract points for inconsistent answers; however, consistency between the free-response and CMR versions was greatest when using the more nuanced rubric \cite{wilcox2014cmr}.  

After initial development of the CMR version of the CUE, we performed a direct comparison of student performance on the CMR and free-response versions with students at University of Colorado Boulder (CU) \cite{wilcox2014cmr}.  No statistically significant differences were observed in students' average score on the two versions.  Moreover, student performance on individual questions was largely consistent between the two versions, and a qualitative analysis of student responses to the free-response version showed a high degree of consistency between the nature of student responses in the two different formats.  Overall, this study found that, for this population of students, the CMR CUE represented an easily-graded assessment that produced scores consistent with that of a free-response instrument \cite{wilcox2014cmr}.  

The goal of this paper is to report on the broader statistical validation of the CMR CUE for independent implementation across a range of student populations.  After a description of the student population, data sources, and analysis (Sec.\ \ref{sec:methods}), we report multiple test statistics relating to the validity and reliability of the CMR CUE (Sec.\ \ref{sec:Vresults}) including: item difficulty; item and whole-test discrimination; internal consistency; and overall consistency of the CMR CUE with other measures of student performance.  We also present a more detailed analysis of student responses to individual questions to investigate consistency across an individual student's responses (Sec.\ \ref{sec:Vresults}) as well as how student responses can be used to gain insight into underlying student difficulties (Sec.\ \ref{sec:Dresults}).  Finally, we end with a discussion of limitations and implications (Sec.\ \ref{sec:discussion}).

\section{\label{sec:methods}Context and Methods}

Following the initial comparison of the CMR and free-response versions of the CUE reported previously \cite{wilcox2014cmr}, we set out to more robustly establish the validity and reliability of the new version as an independent instrument.  To do this, we expanded our data collection with an emphasis on including additional students and instructors at multiple institutions beyond the developing institution.  We recruited instructors to pilot the CMR CUE in several ways including soliciting participants during talks and posters presenting the results of the initial comparison study at professional meetings and workshops (e.g., the American Association of Physic Teachers summer meetings).  The new version was also uploaded to our password protected online materials repository (see \cite{website}) where it can be accessed by any physics instructor interested in using our transformed course materials.  We also contacted a number of colleagues working in PER who facilitated putting us in contact with the instructor in their department who was teaching electrostatics.  

Ultimately, we collected post-test CUE data from 15 courses spanning 9 institutions and 13 instructors.  Institution and course characteristics are shown in Table \ref{tab:institutions}.  We also have pretest data from 13 of these courses.  Pretests were administered in the first week of class as either 20 min in-class activities (N=7) or as an out-of-class online survey (N=6).  For all courses but one, post-tests were administered in the last week of class as a 50 min in-class activity.  In one case, the post-test was given as an out-of-class online survey.  

\begin{table}
\caption{General characteristics of each institution (instit.) where we collected post-test CMR CUE data.  N indicates the number of responses rather than the total number of students enrolled.  \\ * Highest degree offered directly by the Physics Department. \\ **These courses include the 2 semesters in which we conducted the comparison study described in Ref.\ \cite{wilcox2014cmr}.  Only data from students who took the CMR version is included in the total N from these courses.  \\ $^\dagger$ The post-test was taken online at this institution.}\label{tab:institutions}  
\begin{tabular}{ l l l l l l }
  \hline
  \hline
  Instit. \hspace{4mm}& Instit. \hspace{4mm}& Highest \hspace{4mm}& Size & N & N \\
  Code & Type & Degree* & (undergrads) & Courses \hspace{2mm}& Students\\
  \hline
  R1-A & Public & Ph.D. & 25,000 & 5** & 193\\
  R1-C & Public & Ph.D. & 37,000 & 1 & 40 \\
  R1-D & Public & Ph.D. & 40,000 & 1 & 30 \\
  R1-E & Public & Ph.D. & 29,000 & 1$^\dagger$ & 67 \\
  R2-A & Public & Ph.D. & 19,000 & 2 & 33 \\
  BG-B & Private & B.S. & 4,000 & 2 & 23 \\
  BG-C & Private & B.A. & 3,000 & 1 & 8 \\
  BG-E & Private & B.S. & 2,000 & 1 & 8 \\
  BG-F & Private\hspace{2mm} & B.S. & 3,000 & 1 & 19 \\
  \hline
  \hline
\end{tabular}
\end{table}

To what extent in-class implementations of the pre and post-tests can be compared to out-of-class, online implementations is still an open question that we will not attempt to robustly answer here.  However, for students at CU who took both pre and post-test, pretest data show an average score of $31.8 \pm 1.5$\% when the pretest was taken online (N=102) compared to $30.9 \pm 1.5$\% when taken in-class (N=126).  This indicates that, for the pretest, in-class and online implementations are likely comparable.  The post-test, however, is a considerably longer and harder instrument, and it may be that scores on a 50 min assessment administered online and in-class are not directly comparable.  However, for the single course where the post-test was given online, the average score and standard deviation were consistent with that from in-class implementations.  Moreover, the inclusion of this course does not significantly change any of the statistics or conclusions reported in the rest of this section.  As such, we have opted to include these data in the following analysis in order to realize greater statistical power.  

Consistent with the majority of conceptual assessments in physics, our analysis of the validity and reliability of the CMR CUE will be guided by Classical Test Theory (CTT) \cite{engelhardt2009ctt}.  CTT posits a number of characteristics of a high quality assessment and provides various test statistics that quantify how well an instrument matches these characteristics.   For polytomously scored assessments like the CMR CUE, these statistics include \cite{engelhardt2009ctt}: item difficulty as measured by the average score on each individual item, item discrimination as measured by Pearson Correlation Coefficients \cite{cohen2013statistical} of item scores with the rest of the test, internal consistency as measured by Cronbach's Alpha \cite{cortina1993ca}, and whole test discrimination as measured by Ferguson's Delta \cite{ding2009mcanalysis}.  Each of these test statistics will be discussed in greater detail in Sec. \ref{sec:Vresults}.  

One significant drawback of CTT is that all test statistics are population dependent.  As a consequence, there is no guarantee that test statistics calculated for one student population (e.g., physics students at a community college) will hold for another population (e.g., physics students at a university).  For this reason, scores on assessments validated through the use of CTT can only be clearly interpreted inasmuch as the student population matches the population with which the assessment was validated.  For additional discussion of the limitations of CTT, see Ref.\ \cite{wallace2010irt}.  To address the shortcomings of CTT, psychometricians later developed Item Response Theory.  In the simplest IRT model (i.e., the Rasch model \cite{ding2011rasch}), a student's performance on individual items is assumed to depend only on their latent ability and the item difficulty.  More complex IRT models also include parameters to account for item discrimination and student guessing.  For test items that fit this model, all item and student parameters can be determined in such a way as they are independent of both population and test form \cite{ding2009mcanalysis, baker2001irt}.  

Despite the appeal of generating population-independent parameters, there are several significant drawbacks to IRT as a potential tool to develop upper-division physics assessments.  Even the simplest dichotomous IRT models require large N ($>$100) to produce estimates of item and student parameters that are reliable enough for low-stakes testing \cite{deayala2009polyIRT, ding2009mcanalysis}.  This number increases for more complex models that, for example, include item discrimination parameters, or for instruments with polytomous scoring \cite{deayala2009polyIRT}.  The small class sizes typical of upper-division physics would necessitate classroom testing at multiple institutions, possibly over multiple semesters, to collect this volume of data.  Additionally, in order for the parameters generated by IRT to be truly population independent, they must fit the appropriate IRT model.  Crafting a large number of items that fit these models often requires multiple iterations of preliminary testing, further increasing the number of students necessary to develop and validate an assessment.  Due in large part to the logistical barriers to IRT, this analysis will exclusively utilize CTT. 

\section{\label{sec:Vresults}Results: Statistical Validation}

\begin{figure}
    \includegraphics[width=\columnwidth]{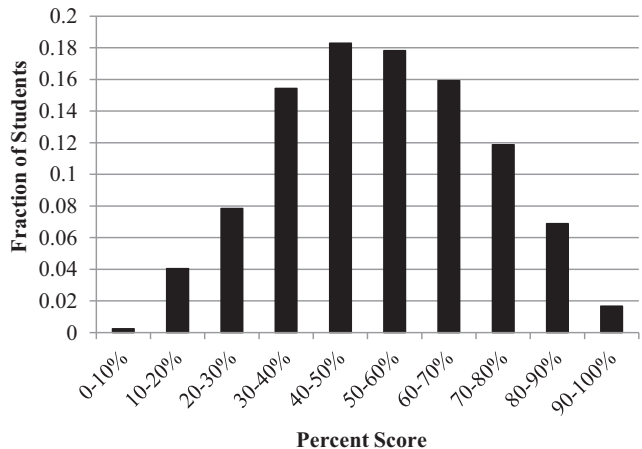}
  \caption{Distributions of scores (N=421) on the CMR CUE from 15 courses at the institutions described in Table \ref{tab:institutions}.  These data did not pass a statistical test for normality (Anderson-Darling test, $p = 0.03$). }\label{fig:fullScoreDist}
\end{figure}

This section presents the statistical validation of the CMR CUE.  Using the nuanced grading rubric described in Sec.\ \ref{sec:introduction}, the overall average on the CMR CUE is $52.6 \pm 0.9$\% when treating students as data points.  The distribution of N=421 scores are shown in Fig.\ \ref{fig:fullScoreDist}.  The distribution is slightly non-normal (Anderson-Darling test \cite{stephens1974ad}, $p=0.03$), due in part to the slight positive skew.  Averaging by students differentially weights the impact of large courses, which, in these data, come exclusively from large research institutions (Table \ref{tab:institutions}).  This effect can be reduced by considering performance by course, rather than by students.  Taking the mean of the average scores for each course, the overall performance on the CMR CUE is $50.3 \pm 2.5$\%.  With only N=15 courses, the difference between the by course and by student averages is not statistically significant; however, we argue that a difference of 2\% is also not of practical significance.  Thus we will treat student scores as data points for the remainder of this analysis.  

\subsection{Criterion Validity}

To establish the extent to which scores on the CMR CUE are consistent with other, related learning outcomes, we would ideally correlate these scores with final course grades and/or aggregate exam scores for all students in our sample.  Unfortunately, we only have access to final course and exam scores for a subset of the students at CU (N=154, 4 courses) and for none of the external institutions.  For this subset of our sample, we find a high correlation of CMR CUE scores with aggregate exam score (Pearson's Correlation Coefficient \cite{cohen2013statistical} $r=0.71$) as well as final course score ($r=0.64$).  To account for differences between the average exam, course, and CUE scores between the semesters, the correlations above are based on standardized scores (z-scores) calculated separately for each class using the class mean and standard deviation.  This finding establishes the criterion validity of the CUE for the student population at CU; however, we are not able to extend this conclusion to external institutions with the available data.  

\subsection{Item Difficulty}

\begin{figure}
    \includegraphics[width=\columnwidth]{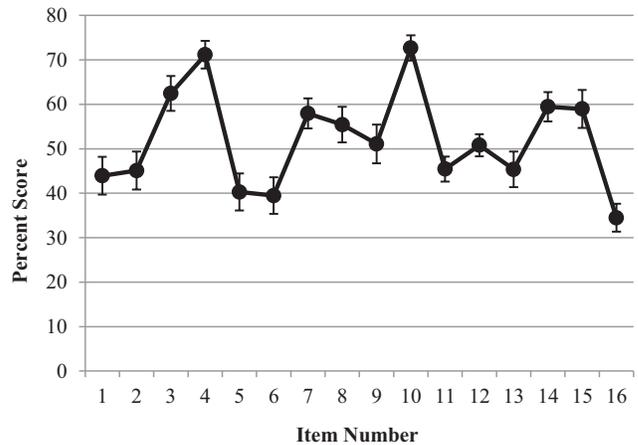}
  \caption{Average scores for each item on the CMR CUE (N=421).  Error bars represent a 95\% confidence interval (double the standard error on the mean).  Score distributions for each individual item are not necessarily normally distributed.  }\label{fig:fullIdifficulty}
\end{figure}

In addition to looking at the overall performance of students on the CMR CUE, we characterize the difficulty of each item by looking at the average score by question (Fig.\ \ref{fig:fullIdifficulty}).  Item difficulties for all questions fall between 30-75\%.  We are not aware of a well-established range of acceptable values for item difficulty on polytomously scored items.  However, for dichotomously scored items where item difficulty is measured as the percent of students who answer each item correctly \cite{ding2006bema}, it is typically argued that ideal values should fall half-way between 100\% and the percent expected by random guessing \cite{doran1980measurement}.  This maximizes the potential discriminatory power of each item.  Since not all items will achieve this ideal, one standard range for acceptable values is 30-90\% \cite{ding2006bema}.  Extending this same logic of maximizing the potential discriminatory power of each item as well as the test as a whole, we argue item difficulties for our polytomously scored items fall within an acceptable range, with no single item being too easy or too hard to contribute to the overall discrimination of the test.  

Scores on each individual item are rarely normally distributed.  This is in part an artifact of the grading scheme in which there are a finite number of potential point combinations (typically between 0-5 pts in 0.5-1 pt intervals).  For this reason, the median score on each item is often different from the average score. 

\subsection{Discrimination}

\begin{figure}
    \includegraphics[width=\columnwidth]{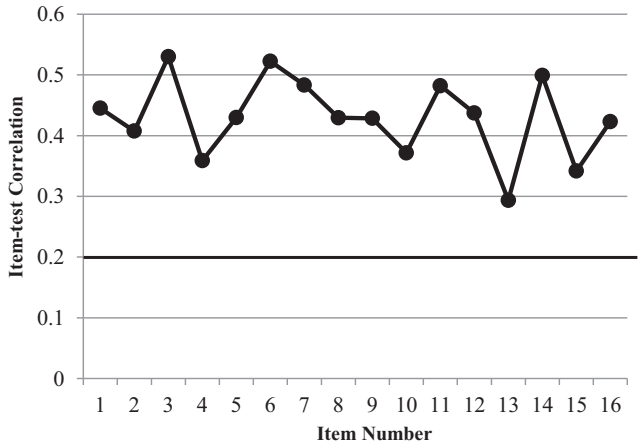}
  \caption{Item-test correlations (as measured by Pearson's $r$) for each of the items on the CMR CUE.  For N=421 any correlation greater than 0.09 is significant at the $p<0.05$ level \cite{weathington2012table}, thus item-test correlations are statistically significant for all items.  The conventional cutoff for an acceptable correlation (0.2) is marked as a bold line.  }\label{fig:fullIdiscrim}
\end{figure}

One preliminary indication of the whole-test discrimination of the CMR CUE comes from the overall spread in the distribution of students' scores (Fig.\ \ref{fig:fullScoreDist}).  These scores span nearly the full range of possible scores (from 0-100\%) with a minimum score of 4.3\% and a maximum score of 92.5\%.  Thus the students are well distributed across the range of possible scores.  As another measure of the whole-test discrimination of the CMR CUE, we use Ferguson's Delta \cite{ding2009mcanalysis}.  Ferguson's Delta is a measure of how well scores are distributed over the full range of possible point values (0-93 points).  For the full population of students, Ferguson's Delta is 0.99.  Delta can take on values between [0,1] and anything above 0.9 indicates good discriminator power \cite{ding2006bema}.  

We also examine the discrimination of each individual item by comparing a student's score on that item to their performance on the rest of the test.  Item-test correlations for all 16 items are shown in Fig.\ \ref{fig:fullIdiscrim}, and all correlation coefficients fall between 0.28-0.55 and are statistically significant given N=421 \cite{weathington2012table}.  As has been done before \cite{chasteen2012cue}, we adopt the standard cutoff of $r=0.2$ used for dichotomously scored items \cite{ding2006bema} to argue that all items on the CMR CUE demonstrate acceptable discriminatory power.  

\subsection{Consistency}

\begin{figure*}
    \includegraphics[width=7in]{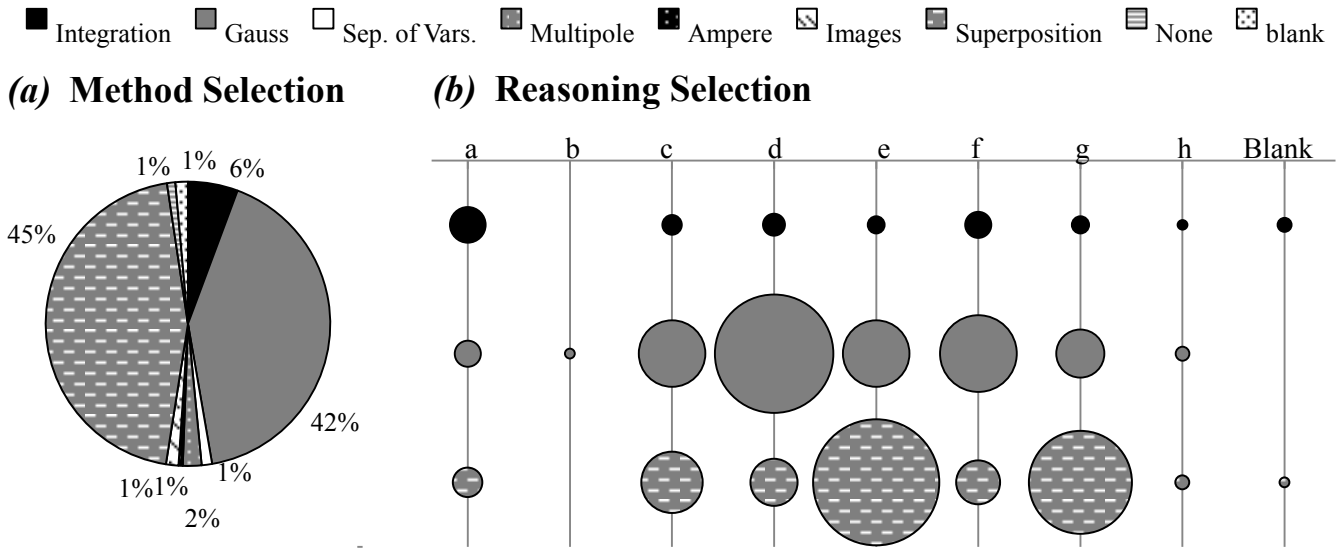}
  \caption{\emph{(a)} Method selections for N=421 students on Q5 (Fig.\ \ref{fig:exampleCMR}). \emph{(b)} Reasoning selections for the subset of students who selected each of the three most common Methods: Direct Integration (N=24), Gauss's Law (N=176), and Superposition (N=191).  Number of responses for each reasoning element is proportional to the area of the circle.  }\label{fig:consistencyQ5}
\end{figure*}

The consistency of scores on individual items or subsets of items is another important property of the CMR CUE.  We utilize Cronbach's Alpha as a conservative measure of the internal consistency of the CMR CUE as a whole.  Cronbach's Alpha can be interpreted as the average correlation of all possible split-half exams \cite{cortina1993ca}.  Alpha is a conservative measure because it assumes a unidimensional assessment, and, while we have no \emph{a priori} reason to assume that the CUE measures a single construct, multidimensionality will tend to drive alpha downwards \cite{cortina1993ca}.  For our population of students, we calculate a value of $\alpha=0.82$.  For a test used for low-stakes testing of individuals rather than just groups, the commonly accepted threshold is $\alpha > 0.8$ \cite{engelhardt2009ctt}.  Thus the CMR CUE demonstrates an acceptable level of internal consistency.  

In terms of the new CMR format, there is another aspect of consistency that is important to consider.  As the name implies, the majority of the questions on the \emph{coupled} multiple-response CUE have several subparts whose scores and/or content are coupled, either explicitly (as with the Method/Reasoning type items, Fig.\ \ref{fig:exampleCMR}) or implicitly (i.e., there is an opportunity for a student to be consistent or inconsistent in their responses to consecutive subparts).  For example, the distribution of method selections for the item shown in Fig.\ \ref{fig:exampleCMR} are given in Fig.\ \ref{fig:consistencyQ5}\emph{(a)}.  The two most common methods are Gauss's Law and Superposition, with Superposition being the correct response.  Fig.\ \ref{fig:consistencyQ5}\emph{(b)} breaks down the reasoning choices for students who selected each of these methods.  There is a clear qualitative difference between the reasoning elements selected by these two sets of students.  Students who chose Superposition were more likely to select reasoning elements `e' and `g', which represent the two elements required to fully justify Superposition as the easiest method.  Alternatively, students who selected Gauss were more likely to select reasoning elements `d' and `f'.  Both of these elements are consistent with the use of Gauss's law and represent the commonly expressed justifications for using Gauss to solve this problem.  

While Fig.\ \ref{fig:consistencyQ5} qualitatively suggests a certain degree of consistency between students' method and reasoning selections, we also wanted to get a more quantitative sense of students' consistency.  To do this, we assigned a consistency code to students' response to each question (excluding Q8, Q11, Q14, \& Q15 which have no consistency check).  Students were coded as `consistent' if they selected at least one of the reasoning elements that supported their specific choice of method/answer and no inconsistent elements.  Alternatively, if they selected any reasoning elements that were directly inconsistent with their choice of method, they were coded as `inconsistent' regardless of whether they also selected some consistent reasoning elements.  The remaining subset of students were coded as `neither', meaning they left one of the two parts blank, chose the `None of These' method option, or selected only reasoning elements that were neither directly consistent nor inconsistent with their choice of method.  For example, on Q5 (Fig.\ \ref{fig:exampleCMR}), the combinations \emph{(Method, Reasoning)=(B,df)} or \emph{(A,a)} would both be coded as `consistent', whereas the combinations \emph{(B,bd)} or \emph{(A,af)}, would be coded as `inconsistent, and the combinations \emph{(B,g)} or \emph{(A,c)} would be coded as `neither.'  

\begin{figure}[b]
    \includegraphics[width=\columnwidth]{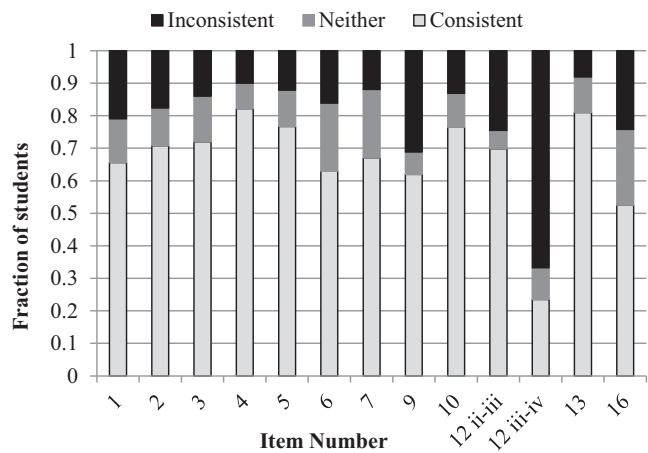}
  \caption{Fraction of students coded as `consistent' (at least one consistent reasoning element and no inconsistent ones), `inconsistent' (any inconsistent reasoning elements), or `neither' (neither consistent nor inconsistent reasoning elements) on each of the questions that have consistency checks.  Note, there are no consistency checks on Q's 8, 11, 14 \& 15, but there are two possible consistency checks in Q12: between subparts ii and iii, and between subparts iii and iv.  }\label{fig:fullConsistency}
\end{figure}

The breakdown of the fraction of students receiving each consistency code is given in Fig.\ \ref{fig:fullConsistency}.  On all questions but one, the fraction of consistent students is $\ge$0.5, and the fraction of inconsistent students is $\le$0.32.  Consistency between Q12 subparts iii and iv is noticeably lower than on other questions.  These two subparts ask for qualitative graphs of $E_z$ and $V$ from a finite disk of charge and, for any given response to subpart iii, there is, at most, one consistent response to subpart iv.  The relatively small number of potential consistent response patterns and the fact that consistency between these subparts is not explicit in the problem statement both contribute to the greater degree of inconsistency on this question.  For all questions, consistent responses do not come exclusively from correct responses.  In other words, many students are consistent even when they are incorrect.  We take this as an indication that the majority of students are connecting their answers and reasoning selections in reasonable and meaningful ways rather than randomly guessing.  This finding further supports the overall validity of the multiple-response format.

\section{\label{sec:Dresults}Accessing Student Difficulties with the CMR CUE}

The previous sections have supported the validity and reliability of the CMR CUE according to Classical Test Theory.  However, in addition to providing a quantitative measure of student outcomes, the CUE also presents an opportunity to gain insight into student difficulties.  For example, we have used student responses to several CMR CUE questions in our investigations of student difficulties with the Dirac delta-function \cite{wilcox2015delta} and separation of variables \cite{wilcox2015thesis}.  In this section, we focus on the question in Fig.\ \ref{fig:exampleCMR} (Q5) as an example of using the CMR CUE to think about student difficulties.  The distributions of student responses to the remaining questions are available from Ref.\ \cite{wilcox2015thesis} but will not be discussed in further detail here.  

Q5 (Fig.\ \ref{fig:exampleCMR}) presents students with a solid sphere with an off-center, spherical cavity carved out of it and asks for the easiest method to find $\vec{E}$ or $V$ outside the sphere.  The correct response is Superposition (`G') because you can treat this situation as two oppositely charged spheres (`e') and superpose the electric fields (`g') from each uniform sphere (`c') individually to determine the total electric field at point P.  It is also possible, though \emph{much} more difficult, to solve this problem through Direct Integration (`A') via Coulomb's Law (`a').  The distribution of method selections from this population of students is shown in Fig.\ \ref{fig:consistencyQ5}\emph{(a)}.  Almost half of the students (45\%, N=191 of 421) correctly selected Superposition as the easiest method, and only a small number (6\%, N=24 of 421) selected the more difficult method, Integration.  Of the remaining students, the overwhelming majority (42\%, N=176 of 421) selected Gauss's Law.  

There are at least two possible reasoning paths that could lead a student to select Gauss's law as the method for this question \cite{zwolak2013CUErubric}.  First, they are imagining using a single large Gaussian sphere centered on the origin of the solid sphere (not the cavity) to calculate $\vec{E}$ from $Q_{enclosed}$ (consistent with reasoning elements `d' and `f').  Alternatively, they could be imagining using two Gaussian spheres, one centered on a solid, uniform sphere and one on a solid, negatively charged sphere in place of the cavity (consistent with reasoning elements `e' and `g'). The latter strategy is correct while the former is fundamentally incorrect.  

To distinguish between these two lines of reasoning, we must examine the reasoning selections of those students who selected Gauss's law (Fig.\ \ref{fig:consistencyQ5}\emph{(b)}).  The two most common reasoning elements selected by these students are `d' and `f', which supports the conclusion that the majority of these students were following the first (incorrect) line of reasoning.  Indeed, of the students who selected Gauss's law, only a tenth (11\%, N=20 of 176) did not select one or both of reasoning elements `d' or `f'.  Only 10 of the remaining students selected both reasoning elements `e' and `g,' suggesting that they were using the second (correct) line of reasoning.  This finding is consistent with previous research \cite{pepper2010gauss} and our own findings \cite{wilcox2013acer} that suggest students often misapply Gauss's law.  In this case, the majority of students have argued that the location of the cavity does not matter, suggesting that either they have not recognized that the asymmetrical location of the cavity breaks the symmetry of the electric field, or that they have not recognized the asymmetry of the electric field eliminates Gauss's law as a potential solution method.  However, it is not possible to decide which of these two issues is at play for a particular student given only their response to this question.  

As Superposition is the correct response to this question, it is tempting to assume that any student selecting method `G' understands the correct solution method.  This conclusion is generally supported by the observation that the most common reasoning elements selected by these students are `e' and `g.'  However, just under a fifth of students who selected Superposition (18\%, N=34 of 191) also selected one or both of reasoning elements `d' and `f', suggesting that these students were thinking only about superposition of charges (i.e., $Q_{large}-Q_{small}$) rather than fields (i.e., $\vec{E}_{large}-\vec{E}_{small}$).  This distinction between superposition of charges rather than fields was also observed in previous research examining student responses to the free-response version of the CUE \cite{zwolak2013CUErubric}.  Both this result and the finding that a small number of students (N=10) selected Gauss's Law along with reasoning elements that suggest a correct strategy, underscore the importance of asking students to express their reasoning to avoid misinterpreting student responses.

\section{\label{sec:discussion}Summary and Discussion}

We previously created a multiple-response version of an existing upper-division conceptual assessment, the CUE.  This new version utilizes a novel approach to multiple-choice questions that allows students to select multiple reasoning elements in order to construct a complete justification for their answers.  By awarding points based on the accuracy and consistency of students' selections, this assessment has the potential to produce scores that represent a more fine-grained measure of students' understanding of electrostatics than a standard multiple-choice test.  Previous research demonstrated that the multiple-response and free-response versions of the CUE resulted in similar student performance and showed a high degree of consistency on multiple measures of test validity and reliability.  

We collected scores on the CMR CUE from multiple courses at multiple institutions.  These data support the validity and reliability of the instrument as measured by various test statistics including item difficulty, item discrimination, and internal consistency.  We also examined the consistency of students' responses on consecutive subparts of individual questions.  These data showed that the majority of students selected responses that were internally consistent even when the overall response was incorrect.  Additionally, as an example of using the CMR CUE to gain insight into student difficulties, we demonstrated that student responses to one question support the findings from previous research that students tend to misapply Gauss's law in non-symmetric situations.  These finding support the overall validity of the CMR CUE as a research-based assessment that can be used to reliably measure some aspects of student learning in upper-division electrostatics.  

Some potential limitations of the CMR CUE include its content coverage and presentation.  Both versions of the CUE were designed to be consistent with Griffiths text \cite{griffiths1999em} in terms of both scope and wording.  Instructors not using Griffiths should carefully examine the content of the CUE to ensure that it matches their content learning goals and coverage.  For example, feedback from some external institutions has indicated that some students may be less likely to use or interpret the term `superposition' in the same way as it is used in the CUE \cite{zwolak2013CUErubric}.  Moreover, the CMR CUE was validated using Classical Test Theory, and thus all test statistics are population dependent.  If used to assess a significantly different population of students, care should be taken to ensure that students' scores can still be reliably and accurately interpreted.

\begin{acknowledgments}
Particular thanks to Stephanie Chasteen for all her work developing and validating the original, free-response version of the CUE, and to the members of the PER@C group their feedback.  This work was funded by NSF-CCLI Grant DUE-1023028 and an NSF Graduate Research Fellowship under Grant No. DGE 1144083.
\end{acknowledgments}

\bibliography{master-refs-04-01-15}
\bibliographystyle{apsper}   % if natbib is available

\end{document}